\documentclass[prd,twocolumn,showpacs,preprintnumbers]{revtex4}
\usepackage{hyperref,amssymb,amsmath,mathrsfs,bm,graphicx}
\newcommand{\be}{\begin{equation}}
\newcommand{\ee}{\end{equation}}
\usepackage{color}
\begin{document}
\title{Central equation of state in spherical characteristic evolutions}
\author{W. Barreto}
\affiliation{Centro de F\'\i sica Fundamental, Facultad de Ciencias,
Universidad de Los Andes, M\'erida, Venezuela}
\author{L. Castillo}
\affiliation{Departamento de F\'\i sica, Escuela de Ciencias, N\'ucleo de Sucre,
Universidad de Oriente, Cuman\'a, Venezuela}
\author{E. Barrios}
\affiliation{Postgrado en F\'\i sica Fundamental, Facultad de Ciencias,
Universidad de Los Andes, M\'erida, Venezuela}
\date{\today} 
\begin{abstract}
We study the evolution of a perfect--fluid sphere coupled to a scalar radiation field. By ensuring a Ricci invariant regularity as a conformally flat spacetime at the central world line
we find that the fluid coupled to the scalar field satisfies the equation of state $\rho_c+3p_c=$ constant at the center of the sphere, where the energy $\rho_c$
density and the pressure $p_c$ do not necessarily contain the scalar field contribution.
The fluid can be interpreted as anisotropic and radiant because of the scalar field, but it becomes perfect and non radiative at the center of the sphere.
These results are being currently considered to build up a numerical relativistic hydrodynamic solver.
\end{abstract}

\maketitle
\section{Introduction}
The simulation of gravitational radiation from collapsing massive sources is one of the main motivations for
numerical relativity \cite{font,winicour}.
In the characteristic formulation Isaacson, Welling, and Winicour \cite{iww82} explored the production of gravitational waves for axially symmetric distributions of matter within the Winicour--Tamburino
framework \cite{tw66}. Some years later, fundamental studies in 1D and 3D \cite{gw92}, \cite{papadopoulos}, \cite{lehner}, were followed by a 3D code able to treat matter by Bishop
et al. \cite{bglmw}.
A relatively simpler system is a spherically symmetric distribution of a perfect fluid coupled with scalar radiation \cite{sfp02}.
For other symmetries other than spherical, the mathematical problem has
a very similar structure \cite{siebel}. 
In these contexts we can take full advantage of the characteristic approach to treat matter and radiation.

We recently reported an unexpected unity in the treatment of matter by the explicit use of
Bondian observers \cite{b64} to describe matter in numerical relativity \cite{barreto}. These observers offer an Eulerian (non--comoving) description (global) with the spirit of Lagrangian observers (local and comoving). 
Consequently, we proposed the post--quasistatic approximation
(PQSA) \cite{hbds02} as a test bed for numerical relativity. 
The PQSA starts from any interior static solution and leads to a system of ordinary differential
equations for quantities evaluated at the boundary surface of the fluid distribution. The numerical solution of this system allows the modeling of self--gravitating spheres
whose static limit is the original ``seed'' solution. The approach is based on the introduction of conveniently defined effective variables and heuristic ansatzs, whose
rationale and justification become intelligible within the
context of the PQSA. In the quasistatic approximation, the effective variables coincide with the physical variables (pressure and density); the method may be regarded as an
iterative method with each consecutive step corresponding to a stronger departure from equilibrium. It has been shown that the effective variables are exactly the conservative and
flux variables in the standard ADM 3 + 1 formulation \cite{barreto}.

Authors commonly refer to the scalar field as a model of matter
distribution that simplifies
the treatment of the hydrodynamic issues. This approach has been
useful to study non linear physics and asymptotic behaviors, especially for central regions
\cite{garfinkle}. Scalar field models have been extended to realistic situations such as gravitational radiation \cite{papadopoulos}.
However, cases where the scalar field is coupled with radiation
require a different approach due to both the confinement effect of matter and the dispersive
nature of radiation. When a scalar field coupled to matter
it can be easily interpreted as an anisotropic fluid.

We develop a  numerical framework to deal with matter coupled to scalar radiation. We perform a detailed study of the central world line at $r=0$, leading to a conformally flat spacetime in that region.
We assume that the radial dependence of the geometrical
and physical variables keep the same dependence as the (quasi or post-quasi) static variables near the coordinate--origin.
Consequently, the central equation of state (CEoS) is unique for all evolutions;
it emerges as a conserved quantity from the field equations.

In what follows we write the field equations for Bondian observers when matter is coupled to scalar radiation, which makes the fluid manifestly anisotropic. The asymptotic approximation to the central world line is discussed with detail in section III, showing that the regular spacetime is conformally flat at $r=0$. Using the PQSA, which belongs to the aforementioned
spacetime, we look the geometrical variables near the coordinate--origin. With all this background
we propose an asymptotic expansion which straightforwardly leads to a CEoS. Some remarks
are finally pointed out.
 
\section{Field equations}
We use the radiation metric in the spherical form \cite{b64}
\begin{equation}
ds^2=e^{2\beta}\Bigl(\frac{V}{r}du^2+2dudr\Bigr)-r^2(d\theta^2+\sin\theta^2 d\phi^2), \label{bm}
\end{equation}
where $\beta=\beta(u,r)$ and $V=V(u,r)$.
In spherical symmetry there exists a well defined notion of quasilocal energy, the Misner--Sharp mass function,
$\tilde m(u,r)$ \cite{ms64} introduced by means of
\begin{equation} 
\tilde m=\frac{1}{2}(r-Ve^{-2\beta}),\label{5} 
\end{equation}
which measures the energy content in the sphere of radius $r$ and it reduces to the
Arnowitt--Deser--Misner and Bondi masses in the appropriate limits.

Now, consider an energy--momentum tensor
for a perfect fluid and a massless scalar field.
We could follow the Tamburino--Winicour formalism \cite{tw66}, in particular
as applied in regular spacetimes where the foliation of light cones emanates from a freely
falling central observer \cite{iww82},\cite{sfp02}. But we choose a different gauge, a Bondian one. For an observer comoving with the fluid, with radial velocity $\omega$ respect to a local Minkowskian observer, the covariant energy--momentum tensor of matter looks like
\begin{equation}
\hat T^M_{\mu\nu}=\left(
\begin{array}{cccc}
\rho & 0 & 0 & 0 \\
0 & p & 0 & 0 \\
0 & 0 & p & 0 \\
0 & 0 & 0 &  p 
\end{array}
\right),
\end{equation}
where $\rho$ and $p$ are the energy density and the pressure respectively.

The energy--momentum tensor for the massless scalar field minimally coupled with gravity 
\begin{equation}
T^\Phi_{\mu\nu}=\nabla_\mu\Phi\nabla_\nu\Phi -\frac{1}{2}
g_{\mu\nu}\nabla^\alpha\Phi\nabla_\alpha\Phi,
\end{equation}
can be characterized as a radiating and anisotropic fluid \cite{hs97}
\begin{equation}
T^\Phi_{\mu \nu}=(\rho^\Phi+p_t^\Phi)u_{\mu}u_{\nu}+\epsilon^\Phi
 l_{\mu}l_{\nu}-p_t^\Phi g_{\mu \nu}+(p^\Phi-p_t^\Phi)
\chi_{\mu}\chi_{\nu},
\end{equation}
with $u^\mu u_\mu=1,$
$l^{\mu}l_{\mu}=0,$
$\chi^\mu \chi_\mu=-1,$
if we identify the four--velocity for an observer at rest in the frame of (\ref{bm}) by 
\begin{equation}
u^\mu=(1-2\tilde m/r)^{-1/2}e^{-2\beta}\delta^\alpha_0, \label{4v}
\end{equation}
and the null and the space--like vectors by
\be l_\mu=(1-2\tilde m/r)^{1/2}e^{2\beta}\delta^0_\mu,\ee
\be\chi_\mu=(1-2\tilde m/r)^{-1/2}\delta^1_\mu.\ee
The scalar energy flux, the scalar energy density, the scalar radial pressure, and
the scalar tangential pressure, respectively are
\be\epsilon^\Phi=e^{-2\beta}[e^{-2\beta}(1-2\tilde m/r)^{-1}\Phi_{,u}^2 -\Phi_{,u}\Phi_{,r}],
\label{rad_phi}\ee

\be\rho^\Phi = p^\Phi = (1-2\tilde m/r)\Phi_{,r}^2/2,\ee

and
\be p_t^\Phi=\Phi_{,u}\Phi_{,r}e^{-2\beta}-p^\Phi,\label{tan_phi}\ee
where the comma denotes partial differentiation respect to the indicated coordinate.
Note that Bondian observers can be purely Lagrangian when 
we deal only with radiation. The observer with four--velocity (\ref{4v}) is resting
at infinity \cite{hbcd07}.

Thus, the Einstein--Klein--Gordon equations can be
written as
\begin{eqnarray}
\frac{\rho+\omega^2 p}{1-\omega^2}+\rho^\Phi + \epsilon^\Phi =
\frac{-\tilde m_{,u}e^{-2\beta}}{4\pi r(r-2\tilde m)}
+ \frac{\tilde m_{,r}}{4\pi r^2}, \label{e1}
\end{eqnarray}

\begin{equation}
\frac{\rho-\omega p}{1+\omega} + \rho^\Phi
=\frac{\tilde m_{,r}}{4\pi r^2},
\label{e2}
\end{equation}

\begin{equation}
\frac{1-\omega}{1+\omega}(\rho+p)+\rho^\Phi+p^\Phi =
 \left(r-2\tilde m\right)\frac{\beta_{,r}}{2\pi r^2},\label{e3}
\end{equation}
\begin{eqnarray}
p+p_t^\Phi&=& -\frac{1}{4\pi}\beta_{,ur}e^{-2\beta}\nonumber\\
&+& \frac{1}{8\pi}\left(1-\frac{2\tilde m}{r}\right) \left(2\beta_{,rr} + 4\beta^2_{,r}- \frac{\beta_{,r}}{r}\right)\nonumber\\
&+& \frac{1}{8\pi r}[3\beta_{,r}(1-2\tilde m_{,r})-
\tilde m_{,rr}], \label{e4}
\end{eqnarray}
and
\begin{equation}
2(r\Phi)_{,ur}=r^{-1}[r e^{2\beta}(r-2\tilde m)\Phi_{,r}]_{,r}, \label{we}
\end{equation}
In absence of matter, $\rho=p=0$, the field
equations reduce to the well known hypersurface equations and wave equation 
for a scalar field \cite{gw92}. The scalar field clearly induces anisotropy
in the matter distribution. Equations (\ref{e1})--(\ref{e4}) are formally the same as those reported many years ago to study anisotropic and radiating matter \cite{chew}.

\section{Analysis of the system near the coordinate--origin}
Some previous investigations consider regularization near $r=0$ \cite{sfp02}, \cite{nc00}, \cite{lake}, \cite{bona-masso}, \cite{alcubierre}. The conditions
for the scalar field, as a matter model, do not necessarily apply to distributions of matter. 
Depending on gauge conditions each procedure to get regular spacetimes may be
cumbersome and tricky, even in vacuum. Initially regular spacetimes can eventually develop singularities \cite{garfinkle}, \cite{choptuik}.
We show a simple way to construct regular spacetimes, near the coordinate--origin, when
the inner spacetime corresponds to a spherical distribution of baryonic matter coupled
to a massless scalar field.

\subsection{Ricci invariant and conformally flat regularity }
Using the Ricci invariant
\begin{equation}
\mathscr{R} = R^{\mu \nu} R_{\mu \nu}=64 \pi^2 \left( \rho^2 + 3 p^2 + \chi \right),
\end{equation}
where 
$$\chi = 2(\rho + p)\left( \frac{1-\omega}{1+\omega}\,\epsilon^{\Phi}
+\frac{1+\omega^2}{1-\omega^2}\,p^{\Phi}_r \right)
+4 \left(p + p^{\Phi}_t\right) p^{\Phi}_t,$$
and the Weyl invariant
\begin{equation}
\mathscr{C} = C^{\mu \nu \alpha \beta} C_{\mu \nu \alpha \beta}=48\left[ \frac{\tilde m}{r^3} -\frac{4}{3}\pi
\left( \rho + p^{\Phi}_t \right) \right]^2
\end{equation}
we can characterize the spacetime near the origin.
The Ricci invariant regularity implies that the physical variables
should be regular at $r=0$. This give us from the field
equations some restrictions on $\tilde m$ and $\beta$, that is,
$\tilde m(u,0)=\tilde m_{,r}(u,0)=\tilde m_{,rr}(u,0)=\beta_{,r}(u,0)=0$; therefore we get $\mathscr{C}|_{r=0}=0$, namely, the
central world line corresponds to a conformally flat spacetime.
We will see below that the post--quasistatic solutions belong to this spacetime class.

\subsection{Static, slow or post--quasistatic motion }
The so called PQSA is well documented  \cite{hbds02}, and its
possible relevance in numerical relativity has been reported recently \cite{barreto}.  For an incompressible--like model and for a polytropic fluid distribution, we find in both cases the following parity in the power expansion near $r=0$ for each geometrical variable: $\tilde m=m_3 r^3+m_5 r^5 + \cdots$ and $\beta=\beta_0 + \beta_2 r^2 + \beta_4 r^4 + \cdots$, where the odd coefficients $m_i$ and the even coefficients $\beta_i$ are constants or very specific functions of time, depending on each case.
This radial dependence is preserved up to the PQSA.

If near $r=0$ the matter's motion is slow,
it seems then reasonable that the physical and geometrical variables preserve the radial dependence.

\subsection{Central equation of state}
We consider the following asymptotic expansions for the metric functions, which correspond
near $r=0$ to
\be
\tilde m=m_3(u) r^3 + \mathcal{O}(r^5), \label{mto0}
\ee
\be
\beta=\beta_0(u)+ \beta_2(u) r^2 +  \mathcal{O}(r^4), \label{betato0}
\ee
and for the escalar field
\be
\Phi=\Phi_0(u) + \Phi_1(u) r + \Phi_2(u) r^2  + \mathcal{O}(r^3). \label{Phito0}
\ee
Now, using the field equations we get the physical variable expansions as a function of $r$ as shown in Table I. 
$\;$
\begin{table}[!ht]
\begin{ruledtabular}
\begin{tabular}{|l|c|c|c|}
$\mathcal{O}$ &$\pi\rho$ (energy density)&$\pi p$ (pressure)&$\omega$\\
\hline
$0$&$\frac{3}{4}m_3-\frac{1}{2}\pi\Phi^2_1$&$\beta_2-\frac{3}{4}m_3-\frac{1}{2}\pi\Phi^2_1$&$0$\\
\hline
$1$  & $\omega_1(\beta_2-\pi\Phi^2_1)+\pi\Phi_1\Phi_2$  & $\omega_1(\beta_2-\pi\Phi^2_1)-2\pi\Phi_1\Phi_2$  &  $\omega_1$        \\
\end{tabular}
\end{ruledtabular}
\caption{Coefficients for the physical variable expansions as a power of $r$ near the coordinate--origin.}
\label{tab:expansion}
\end{table}

From the evolution equation for the scalar field (\ref{we}) we get
\be
\dot \Phi_0 = e^{2\beta_0}\Phi_1, \label{dotPhi0}
\ee
and
\be
\dot \Phi_1 = \frac{3}{2} e^{2\beta_0}\Phi_2. \label{dotPhi1}
\ee

From the field equation (\ref{e4})
\be
\dot\beta_2=[\pi\Phi_1\Phi_2-2\omega_1(\beta_2-\pi\Phi^2_1)] e^{2\beta_0},  \label{dotbeta}
\ee
and from (\ref{e1})
\be
\dot m_3=2\dot\beta_2,\label{dotem}
\ee
where the dot indicates time differentiation. 
This last equation, together with expansions showed in Table I, 
is readily integrated to obtain
\be
\rho_c+3p_c =\text{constant} \label{ee_0},
\ee
where $\rho_c=\rho_0+\rho_0^\Phi$  and $p_c=p_0+p_0^\Phi$.
 
This CEoS is general if the proposed expansions of the metric variables and the scalar field near the origin, in powers of the radial coordinate, is general as well. It is easily verifiable that lesser order expansions are not general enough for the used metric; a higher order does not give additional information. The same is not true for the scalar field. This type of expansion has been used in the construction of spacetimes that eventually develop critical collapse \cite{garfinkle}. 
Our main motivation to specify these expansions was the behavior of
the metric functions in the case of PQSA in radiation coordinates \cite{hjr80}. We supposed
that near $r=0$ all the variables (geometrical and physical) preserve their radial
dependence, even with an embedded scalar field. In fact, the scalar field does not affect
the Ricci invariant spacetime regularity which is conformally flat.
An interesting property is that for eventual maximum values of the central
energy density, the central pressure will have to reach minimum
(possibly negative) values in order to preserve  the quantity $\rho_c+3p_c$.
Also, according to expressions (\ref{rad_phi}) and (\ref{tan_phi}) the fluid is perfect ($p^\Phi_{t0}=p^\Phi_0$) and non radiating ($\epsilon^\Phi_0=0$) at $r=0$; this result is in agreement with our expectations. 
The referential value of the constant in (\ref{ee_0}) corresponds to the static situation ($\omega=0$), limited physically by $-1<\omega<1$. The initial values for $m_3$ and $\beta_2$ can be determined by the initial--central value of the energy--density $\rho_0(0)$. 
We have to specify the initial conditions for $\beta_0$ and $\omega_1$ to integrate the system
of two equations for $\beta_2$ and $m_3$.
For $\omega_1$ we could fit a quadratic polynomial and update at any time. 
Our gauge leads to $\beta_0\ne 0$. Initial values for $\Phi_0$, $\Phi_1$ and $\Phi_2$ will be given by fitting the initial radial profile for the scalar field. For the evolution we have to fit only $\Phi_2$, to evolve $\Phi_0$ and $\Phi_1$ using equations (\ref{dotPhi0}) and (\ref{dotPhi1}).

\subsection{CEoS as the strong energy condition}
It is well known that the strong energy condition reads \cite{wald}
\begin{equation}
\left(T_{\mu \nu}-\frac{1}{2}g_{\mu\nu}T\right) U^{\mu} U^{\nu} \geq 0
\end{equation}
where $U^\mu$ is the Bondian four--velocity for matter
\begin{eqnarray}
U^\mu&=&e^{-2\beta}(1-2\tilde m/r)^{-1/2}\left(\frac{1-\omega}{1+\omega}\right)^{1/2}\delta^\mu_0\nonumber\\
&+& \omega\left(\frac{1-2\tilde m/r}{1-\omega^2}\right)^{1/2}\delta^\mu_1.
\end{eqnarray}
For the present case
\begin{eqnarray}
R_{\mu\nu} U^{\mu} U^{\nu} =-4\pi\left[ (\rho + 3\,p) 
+2 \left(U^{\alpha} \Phi_{,\alpha}\right)^2 \right]\leq 0,
\end{eqnarray}
which can be written explicitly at $r=0$ as
\begin{equation}
R_{\mu\nu} U^{\mu} U^{\nu}|_{r=0}=-4\pi \left( \rho_c  + 3 p_c \right),
\end{equation}
that is, $\rho_c  + 3 p_c \geq 0,$
which means that the CEoS satisfies easily this restriction.

\section{Concluding remarks}
We construct a regular spacetime which is conformally flat at $r=0$; as a realization of this
class of spacetime we have the post--quasistatic solutions, which preserve with evolution the radial dependence
for the inner regions. Doing a PQSA--like treatment near the origin we find a CEoS as a conserved quantity for any distribution of spherical perfect fluid matter coupled with scalar radiation. 

We expect to use the CEoS as a convenient test to check the evolution where profuse radiation could be present. 
Our disclosure of a CEoS was likely possible due to the characteristic formulation. Moreover, Bondian observers makes it quite easy. 
Its use for  the central zone, the boundary surface, and the infinity is leading us to a template code which we are reporting elsewhere.
We wonder if the Bondi--Sachs
gauge \cite{gf07} could lead to a CEoS in ADM 3+1 formulation. 

\acknowledgments  We are grateful to Adriana \'Alvarez, Loren Lockwood, Beltr\'an Rodr\'\i guez--Mueller, Luis Rosales and Carlos Pe\-ralta for their valuable comments.
\end{document}